\documentclass[%
 reprint,
 superscriptaddress,
nofootinbib,
 amsmath,amssymb,
 aip,
]{revtex4-2}

\usepackage{graphicx}% Include figure files
\usepackage{dcolumn}% Align table columns on decimal point
\usepackage{bm}% bold math
\usepackage{hyperref}% add hypertext capabilities
\usepackage{physics}
\usepackage{float}
\usepackage{xcolor}
\newcommand{\ld}[1]{\textcolor{black}{#1}}

\usepackage{balance}

\begin{document}

\preprint{APS/123-QED}

\title{Deciphering the Small-Angle Scattering of Polydisperse Hard Spheres using Deep Learning}% Force line breaks with \\
\author{Lijie Ding}
\affiliation{Neutron Scattering Division, Oak Ridge National Laboratory, Oak Ridge, TN 37831, USA}

\author{Changwoo Do}
\email{doc1@ornl.gov}
\affiliation{Neutron Scattering Division, Oak Ridge National Laboratory, Oak Ridge, TN 37831, USA}

\date{\today}% It is always \today, today,

\begin{abstract}
We introduce a deep learning approach for analyzing the scattering function of the polydisperse hard spheres system. We use a variational autoencoder-based neural network to learn the bidirectional mapping between the scattering function and the system parameters including the volume fraction and polydispersity. Such that the trained model serves both as a generator that produce scattering function from the system parameters, and an inferrer that extract system parameters from the scattering function. We first generate a scattering dataset by carrying out molecular dynamics simulation of the polydisperse hard spheres modeled by the truncated-shifted Lennard-Jones model, then analyze the scattering function dataset using singular value decomposition to confirm the feasibility of dimensional compression. Then we split the dataset into training and testing set and train our neural network on the training set only. Our generator model produce scattering function with significant higher accuracy comparing to the traditional Percus-Yevick approximation and $\beta$ correction, and the inferrer model can extract the volume fraction and polydispersity with much higher accuracy than traditional model functions. 
\end{abstract}

\maketitle

%\tableofcontents

\section{Introduction}
% significance of solving polydispersity in scattering, falure of traditional approach
Understanding the structural properties of colloidal dispersions\cite{russel1991colloidal} is crucial in fields such as soft matter, materials science, and biology. Techniques like small-angle scattering\cite{lindner2002neutrons} (SAS), including neutron scattering\cite{shibayama2011small,chen1986small} and X-ray scattering\cite{chu2001small,debye1947molecular,schartl2007light}, are indispensable tools for probing these structure at nanoscale. These methods provide scattering functions that encode information about the size, shape, and interactions of particles within the suspension. However, interpreting these scattering functions, especially for complex systems with polydispersity where the particles are of different size, remains a significant challenge\cite{mittelbach1998direct,huang2023model,weyerich1999small}. In such systems, the contributions from the individual particles shapes (form factor) and their arrangement (structure factor) become intrinsically coupled and highly dependent on the size distribution and density, making the direct modeling of these effects analytically complex and often intractable with traditional approaches. Among the polydisperse systems, polydisperse hard spheres\cite{salgi1993polydispersity,wagner1994viscosity} serve as the simplest, yet non-trivial and fundamentally important reference system for understanding the statistical mechanics and physics of dense colloidal suspensions\cite{eckert2022sedimentation,pusey1986phase}. Their well-defined, purely repulsive interactions make them an ideal model for developing and validating new analytical and computational methods for complex scattering problems.
 
% problem with traditional approach
Despite the fundamental importance, accurately modeling the scattering function of polydisperse hard spheres, especially at higher concentrations and polydispersity, has been difficult using conventional analytical theories. Traditional approaches, such as the Percus-Yevick approximation\cite{percus1958analysis,wertheim1963exact} for the structure factor combined with a form factor averaging scheme\cite{chen1986small} ($\beta$ correction for polydispersity), attempt to decouple the particle shape and inter-particle correlations. While these methods offer valuable insights for dilute or monodisperse systems, their fundamental assumptions break down under conditions of increasing volume fraction and polydispersity\cite{katzav2019random,salgi1993polydispersity,zaccone2022explicit,anzivino2023estimating}. This oversight leads to significant discrepancies between theoretical solutions and experimental or simulation data, limiting their predictive power for scattering data analysis. Consequently, there is a pressing need for more robust and accurate methods that can capture the complex scattering behavior of polydisperse systems.

% development of ML lead us to attach this problem using DL
Machine learning (ML)\cite{murphy2012machine,carleo2019machine}, and deep learning\cite{goodfellow2016deep,lecun2015deep}, has transformed data analysis in science by uncovering complex pattern in large datasets. In soft matter physics and scattering analysis, ML models have been applied to analyze scattering function of systems like colloids\cite{chang2022machine,tung2024inferring, ding2025colloids}, lamellar\cite{tung2025scattering,tung2025insights}, and polymer\cite{ding2024mechanical,ding2025charge,ding2025ladder,ding2025deciphering} systems. These ML models utilize simulation data to learn the mapping between the scattering function and the corresponding system parameters. This data-driven, simulation-based approach\cite{cranmer2020frontier,chang2022machine} offers a powerful alternative to traditional analytical method, motivating their use for interpreting the scattering data of interacting polydisperse colloids.

% What we do in this work
In this work, we present a deep learning framework for the bidirectional analysis of scattering functions from polydisperse hard spheres. Leveraging a variational autoencoder\cite{kingma2013auto,doersch2016tutorial} (VAE)-based neural network, we establish a robust and accurate mapping between the scattering function and system parameters: the volume fraction $\eta$ and polydispersity $\sigma$. Our framework specifically comprises a generator model for predicting scattering functions from given $\eta$ and $\sigma$, and an inferrer model for rapidly and accurately extracting these parameters directly from scattering data. We first generate a comprehensive dataset of scattering functions using molecular dynamics\cite{allen2004introduction} (MD) simulations of polydisperse hard spheres, which serves as the ground truth for training and validating our models. We demonstrate that our deep learning generator significantly outperforms traditional analytical approximations, and our inferrer model exhibits high precision in extracting the system parameters, showcasing its potential for efficient inverse analysis. We validate the versatility of our approach by applying it to systems with various particle size distributions, including uniform, normal, and lognormal distribution.

\section{Method}

\subsection{Molecular Dynamics}
To understand the structure of the polydisperse hard spheres, we carry out MD simulations using LAMMPS\cite{thompson2022lammps}. To model the hard sphere potential, we use truncated-shifted Lennard-Jones\cite{lennard1931cohesion,chang2022machine} (LJ) model with
\begin{equation}
    V_{LJ}(r_{ij}) = \begin{cases}
        4\epsilon_{LJ}\left[ \left( \frac{\sigma_{ij}}{r_{ij}}\right)^{12} - \left( \frac{\sigma_{ij}}{r_{ij}}\right)^{6} - \frac{1}{4} \right] & r_{ij} < 2^{\frac{1}{6}}\sigma_{ij} \\
        0 & \textit{otherwise}
    \end{cases}             
\end{equation}
where $\epsilon_{LJ}=100$, $r_{ij}$ is the distance between particle $i$ and $j$, $\sigma_{ij} = 2^{-\frac{1}{6}}(D_i+D_j)/2$. The diameter $D_i$ of each particle is sample from target distribution, and a total of $23,328$ particles are included in the simulation. We perform the Canonical (NVT) ensemble simulation with the temperature $T=1.0$ maintained using Nose-Hoover thermostat\cite{nose1984unified,hoover1985canonical}. Similar approximation has been used in the simulation of hard sphere Yukawa particles\cite{chang2022machine}.

To model the polydispersity of the hard spheres, we consider three kinds of size distribution that commonly used and available in existing software packages such as SASView, \ld{which is widely adopted by the small angle scattering community}: uniform distribution, normal distribution, and lognormal distribution. We use $D_0$ to indicate the mean diameter for the uniform and normal distribution, and the median diameter of the lognormal distribution, $D_0=1$ is also chosen as the natural unit in the simulation.

% talk about volume fraction to number density convention here?

\subsection{Small-angle scattering analysis}
\begin{figure}[!t]
    \centering
    \includegraphics[width=\linewidth]{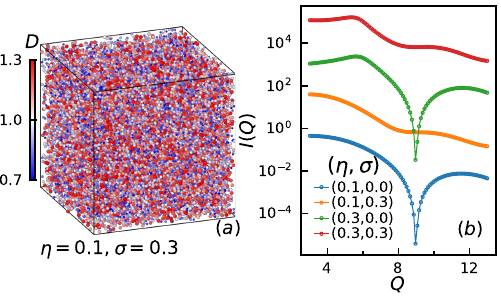}
    \caption{Scattering function of polydisperse hard sphere with uniform distributed size. (a) Snapshot of the system configuration with $\eta=0.1$ and $\sigma=0.3$, such that the diameter of the hard spheres distribute uniformly within $[0.7, 1.3]$. (b) Scattering function $I(Q)$ with different volume fraction $\eta$ and polydispersity parameter $\sigma$. The $I(Q)$ are shifted vertically for better visualization.}
    \label{fig:Iq_and_config}
\end{figure}

Due to the polydispersity, the structure factor of the system and form factor of individual particles are coupled, thus we calculate the normalized scattering function of the polydisperse hard spheres using\cite{lindner2002neutrons,chen1986small}:
\begin{equation}
    I(\vb{Q}) = \left< \frac{\sum_i e^{-i\vb{Q}\vb{r}_i} F(Q;D_i) \sum_i e^{i\vb{Q}\vb{r}_i} F(Q;D_i)}{ \sum_i(\frac{\pi D^3_i}{6})^2}  \right>
    \label{eq:IQ}
\end{equation}
where $\left<\dots\right>$ is the ensemble average of all configurations, $\vb{Q}$ is the scattering vector, $\vb{r}_i$ is the position of particle $i$, $D_i$ is the diameter of particle $i$, and $F(Q;D_i)$ is the form factor amplitude of the hard sphere, which is given by\cite{guinier1956small}:
\begin{equation}
    F(Q;D) = \frac{\pi D^3}{6}\frac{3\left[\sin(Q\frac{D}{2})- Q\frac{D}{2}\cos(Q\frac{D}{2})\right]}{(Q\frac{D}{2})^3}
    \label{eq:FQ}
\end{equation}

Fig.~\ref{fig:Iq_and_config} shows an illustrative configuration of polydisperse hard spheres and examples of scattering function $I(Q)$. Traditionally, scattering intensities for monodisperse hard spheres could be modeled via decoupling structure factor and form factor, where the structure factors are often solutions from the Ornstein–Zernike equation\cite{ornstein1914accidental} using various closure relations such as the Percus-Yevick approximations \cite{percus1958analysis,wertheim1963exact,kinning1984hard} and the intensity can be written as: 
\begin{equation}
    I_{PY}(Q) = S_{PY}(Q)P(Q)
    \label{eq:IPY}
\end{equation}
in which $S_{PY}(Q)$ is the Percus-Yevick approximation of the hard spheres structure factor which depend on the volume fraction $\eta$, and $P(Q)$ is the form factor given by $P(Q) = \left<  F(Q;D)^2 \right>_D/\left<(\frac{\pi D^3}{6})^2\right>_D$. To account for polydispersity of hard spheres, a $\beta$ correction \cite{chen1986small} was introduced, such that
\begin{equation}
    I_{PY\beta}(Q) = (1 + \beta (S_{PY}(Q)-1))P(Q)
\end{equation}
where $\beta = \left<F(Q;D)\right>_D^2/\left<F(Q;D)^2\right>_D$. The detailed calculation of $S_{PY}(Q)$ can be found in the corresponding references\cite{percus1958analysis,wertheim1963exact,kinning1984hard}. Apparently, the traditional method oversees the coupling between the form factor and inter-particle structure factor, and we expect such approximation to fail at higher volume fraction where the interaction between the particles start to reflect on the scattering function. \ld{It is crucial to note that the scattering analysis methods described here are not applicable to optical or light scattering methods\cite{SCHEFFOLD2025581}, where different physical principles and near-field effects need to be considered.}

\subsection{Variational Autoencoder-based neural network}
\begin{figure*}[!t]
    \centering
    \includegraphics[width=\linewidth]{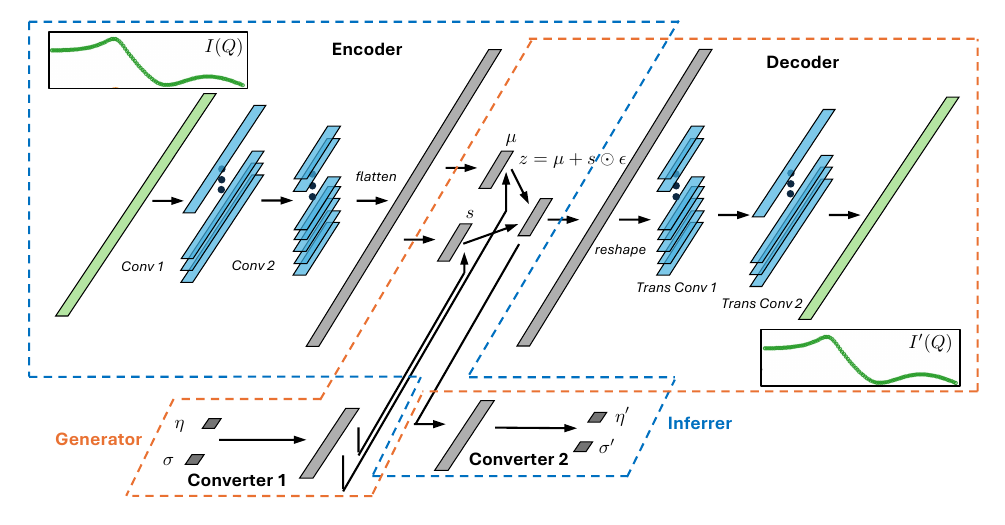}
    \caption{Architecture of the neural network for mapping the scattering function $I(Q)$ with system parameters $(\eta,\sigma)$. The neural network has four main parts: an encoder that consist of 2 convolutional layers that compress the input scattering function to the latent variable; an decoder that decompress the latent variable back to the scattering function using the transposed convolutional layers; converter 1 that map the system parameters to the latent variables; and converter 2 that map the latent variables to the system parameters.}
    \label{fig:NN_architecture}
\end{figure*}

To address the coupling of the structure factor and form factor of the scattering function, and build a neural-network enabled framework for generating scattering function $I(Q)$ and extracting system parameter including volume fraction $\eta$ and polydispersity parameter $\sigma$, we use a VAE to learn the latent representation of the scattering function, and then map the system parameter and the latent variables. 

The whole neural-network architecture is shown in Fig.~\ref{fig:NN_architecture}. The VAE\cite{kingma2013auto} is made of three main components: an encoder, a latent space, and a decoder. The encoder will compress the input scattering function $I(Q)$, a 100 dimensional vector in to the 3 dimensional latent space, and the encoder will reconstruct the scattering function $I'(Q)$ from the latent space. The encoder is consist of two convolutional layers, each with kernel size $9$ and strides 2. The first one has 30 channels, and the second one has 60 channels. After these two convolutional layer, the $I(Q)$ is transformed to a $60\times 25=1500$ dimensional vector. which is then transformed to the latent mean variable $\mu$ and standard deviation variable $s$. These two are then reparameterized into $z=\mu + s\odot \epsilon$ through the normally distributed variable $\epsilon$. The latent variable $z$ is then transform to the output scattering function $I'(Q;\epsilon)$ through two transposed convolutional layers in the decoder, symmetric to which in the encoder. Averaging 100 randomly sampled $\epsilon$, we get the reconstructed scattering function $I'(Q) = \left< I'(Q;\epsilon) \right>_\epsilon$. To train the VAE, we use the loss function:
\begin{equation}
    L_{VAE} =\frac{1}{N} \sum_{I(Q)} \left<\left[\log_{10}I(Q) - \log_{10}I'(Q) \right]^2 \right>_{Q}
    \label{equ:loss_VAE}
\end{equation}
that calculate the Euclidean distance between the log of input and output scattering function, and $N$ is the number of samples.

Moreover, we connect the volume fraction $\eta$ and polydispersity $\sigma$ with latent variable using two converter networks consisted of simple linear layers of dimension 9. As shown in Fig.~\ref{fig:NN_architecture}, the converter 1 maps the system parameter $(\eta,\sigma)$ to the latent variable $(\mu,s)$, and the converter 2 maps the latent variable $z$ to the output system parameter $(\eta',\sigma')$. Combining the converter 1 and the decoder, we get a generator that produces the scattering function from system parameters, while combining the encoder and converter 2, we get the inferrer that directly extract the system parameters from the scattering function. To train the generator, we use the loss function $L_{VAE}$ and firstly freeze the decoder to train the converter 1 only, then, we release the decoder and fine tune the entire generator. As for the inferrer, we introduce a new loss function that calculate the mean square distance between the input $(\eta,\sigma)$ and output $(\eta',\sigma')$:
\begin{equation}
    L_{CVT} = \frac{\sum_{(\eta,\sigma)} \left[(\eta-\eta')^2+(\sigma-\sigma')^2\right]^2}{N}
\end{equation}
Similarly, we firstly freeze the encoder to train the converter 2 only, then fine tune the inferrer with the encoder released. In practice, the neural network is implemented using PyTorch\cite{paszke2019pytorch} and trained using Adam optimizer\cite{kingma2014adam} with CosineAnnealingLR scheduler\cite{loshchilov2016sgdr}. We train the VAE for 1000 epoch, each converter for 300 epoch, and another 200 epoch for fine tuning.

\section{Results}
We start with investigating the dependency of the scattering function on the system parameters including volume fraction and polydispersity of the hard spheres, then train the neural network model to obtain the generator and inferrer for the mapping between system parameters and the scattering function. Next, we demonstrate the accuracy improvement for the scattering function generation using the generator model, comparing with the traditional methods including Percus-Yevick approximation and $\beta$ correction. Later we apply the inferrer to extract system parameters from the scattering function. Without losing generosity, the above results are showcased using the uniform distributed system. Finally, we show the application of our approach on the normal and lognormal distributed hard spheres system.

\subsection{Scattering function of hard spheres}
\begin{figure}[!t]
    \centering
    \includegraphics[width=\linewidth]{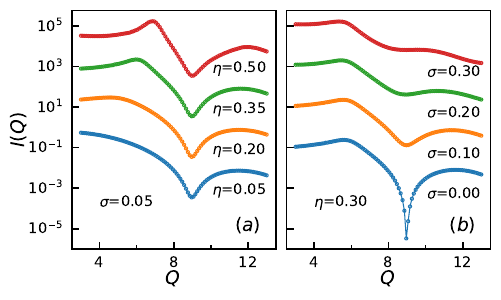}
    \caption{Examples of the scattering function $I(Q)$ for polydisperse hard spheres with uniform distributed diameter size $D\in U(1-\sigma,1+\sigma)$, for various volume fraction $\eta$ and polydispersity $\sigma$. (a) Variation of scattering function for different volume fraction $\eta$ with polydispersity $\sigma=0.05$. (b) Variation of scattering function for different polydispersity $\sigma=$ with volume fraction $\eta=0.3$.}
    \label{fig:Iq_example}
\end{figure}

We first inspect effect of the volume fraction $\eta$ and polydispersity $\sigma$ on the scattering function of the simulated polydisperse hard spheres system with uniformly distributed diameter $D\in U(1-\sigma,1+\sigma)$, where $U(a,b)$ is the uniform distribution in the interval $[a,b]$. As shown in Fig.\ref{fig:Iq_example}, the $\eta$ and $\sigma$ mostly control the shape of the $I(Q)$ at different $Q$ range. The volume fraction $\eta$ mainly affect the interaction peak at low $Q$ while the polydispersity $\sigma$ is mainly responsible for the minimum at middle $Q$. This provides hint for inverse mapping from the scattering function back to the system parameters including volume fraction $\eta$ and polydispersity $\sigma$.

\begin{figure}[!t]
    \centering
    \includegraphics[width=\linewidth]{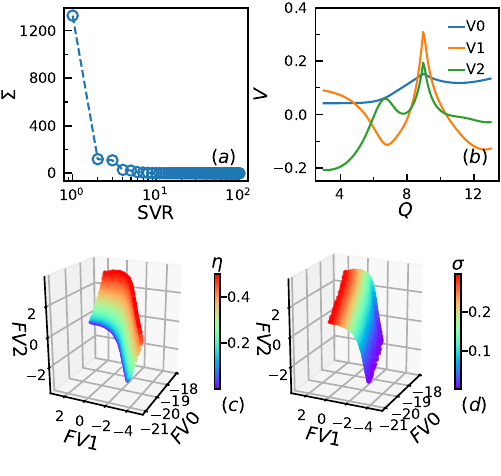}
    \caption{Principal component analysis of the scattering function dataset for $\vb{F} = \{\log_{10}{I(Q)}\}$ of uniform-distribution polydisperse hard spheres. (a) Decay of the singular value entry in $\Sigma$ in $\vb{F}=U\Sigma V^T$ versus the singular value rank (SVR). (b) First 3 singular vectors $(V0,V1,V2)$ of the dataset. (c) Distribution of the volume fraction $\eta$ in the singular value space $(FV0,FV1,FV2)$ in which each $\log_{10}(I(Q))\in \vb{F}$ is projected to the $(V0,V1,V2)$. (d) Distribution of the polydispersity $\sigma$ in the singular value space.}
    \label{fig:svd_combined}
\end{figure}

We then prepare the a dataset consisting of $5,000$ $(\eta,\sigma)$ and corresponding scattering function $I(Q)$ by running the MD simulation for the uniform-distribution polydisperse hard spheres system. The volume fraction and polydispersity are randomly selected from $\eta\in U(0,0.5)$ and $\sigma\in U(0,0.3)$, independently. And the scattering function $I(Q)$ are calculated for structure vector $Q$ uniformly placed on the $100$ grid points within $Q\in[3,13]$. \ld{The $Q$ range is chosen such that we focus on the interaction peak at low $Q$ and also cover the first first dip of sphere form factor. High order minima are not taken into consideration since there is no interaction effect at these $Q$ range.} Combining all of the scattering function into $\vb{F} = \{\log_{10}{I(Q)}\}$, which is a $5000\times 100$ matrix. To analyze the dataset $\vb{F}$, we carry out principle component analysis\cite{jolliffe2016principal} by calculating the singular value decomposition\cite{golub1965calculating} of the dataset $\vb{F}=\vb{U}\vb{\Sigma} \vb{V}^T$, such that $\vb{U}$ is $5000\times 5000$, $\vb{\Sigma}$ is $5000\times 100$, $\vb{V}$ is $100\times100$, and the diagonal entries of $\vb{\Sigma}$ are the singular values of $\vb{F}$, determining the coefficient of the projection of $\vb{F}$ onto $\vb{V}$.

Fig.~\ref{fig:svd_combined}(a) shows the rapid decay of the singular value with its rank, suggesting the feasibility to compress the 100-dimensional vector of scattering function $I(Q)$ into a lower dimensional representation. Fig.~\ref{fig:svd_combined}(b) shows the first three singular vector from $\vb{V}$. Moreover, by projecting the $\vb{F}=\{\log_{10}{I(Q)}\}$ onto the singular vector space spanned by $(V0,V1,V2)$, each scattering function $I(Q)$ can be approximated by the coordinate in the three dimensional space as $(FV0, FV1, FV2)$. Correspondingly, by plotting the distribution of the system parameters $(\eta,\sigma)$ in the $(FV0, FV1, FV2)$ space, as shown in Fig.~\ref{fig:svd_combined}(c) and (d), we can inspect the feasibility to distinguish the $(\eta,\sigma)$ directly from the scattering function. The continuous variation of the system parameters in these plots confirms such possibility for achieving the inverse mapping from the scattering function $I(Q)$ back to the system parameters $(\eta,\sigma)$.

\subsection{Generation of scattering function}
\begin{figure}[!t]
    \centering
    \includegraphics[width=\linewidth]{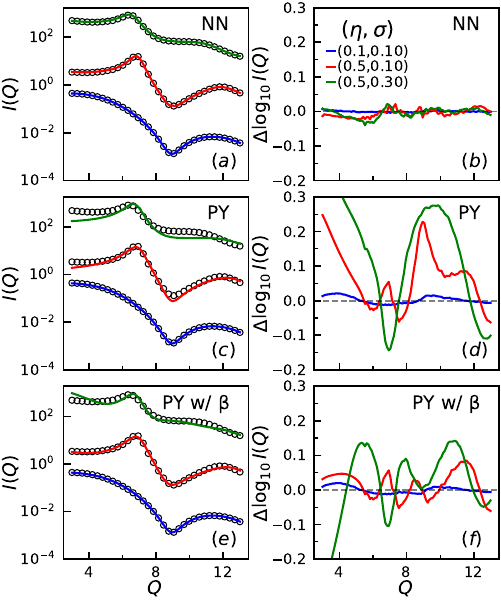}
    \caption{Comparison between the simulation calculated scattering function $I(Q)$ and the corresponding ones generated $I'(Q)$ using different model. All subplots share the same legend as shown in (b).(a) Comparison between 3 example simulation calculated $I(Q)$ of various $(\eta,\sigma)$ and the Neural Network-generated ones. (b) Different between the $I(Q)$ and generated $I'(Q)$, such that $\Delta \log_{10}{I(Q)} = \log_{10}{I(Q)/I'(Q)}$. (c) and (d) Similarly for the Percus-Yevick approximation. (e) and (f) Percus-Yevick approximation with $\beta$ correction. }
    \label{fig:gen_vs_PY_Iq_comparison.pdf}
\end{figure}
We then carry on to train our neural-network model to achieve the bidirectional mapping between the scattering function $I(Q)$ and the system parameters including the volume fraction $\eta$ and polydispersity $\sigma$. Firstly, we randomly split the entire dataset $\vb{F}=\{\log_{10}{I(Q)}\}$ into two parts: training set $\{\log_{10}{I(Q)}\}_{train}$ consist of $4000$ $I(Q)$ and testing set consisting of the rest $1000$ ones. We train the neural network using the training set only. We firstly inspect the trained generator which combine the converter 1 network mapping the $(\eta,\sigma)$ to the latent variables and the decoder network mapping the latent variables to the generated scattering function $I'(Q)$.

To compare our neural-network model with the traditional methods, we benchmark these model-generated scattering function $I'(Q)$ against the simulation calculated $I(Q)$. Fig.\ref{fig:gen_vs_PY_Iq_comparison.pdf} shows few example $I(Q)$ along with the corresponding model-generated $I'(Q)$ from three different methods: neural network generated, Percus-Yevick approximation, and Percus-Yevick approximation with $\beta$ correction. At low volume fraction and low polydispersity, all three models agree well with the simulation results. When the volume fraction and polydispersity increase, our generative model still produce low error as shown in Fig.\ref{fig:gen_vs_PY_Iq_comparison.pdf}(b), whereas the Percus-Yevick approximation has significantly deviate from the simulation data even with the $\beta$ correction applied, as indicated by Fig.\ref{fig:gen_vs_PY_Iq_comparison.pdf}(d) and (f).

\begin{figure}[!t]
    \centering
    \includegraphics[width=\linewidth]{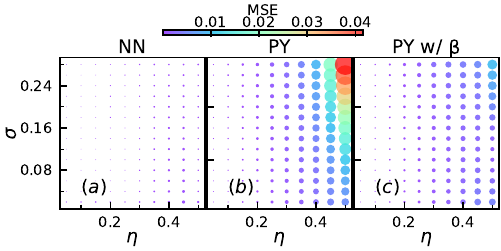}
    \caption{Mean square error $MSE=\left< (\log_{10}{I(Q)/I'(Q)})^2\right>_Q$ of the generated $I'(Q)$ using 3 different models for various volume fraction $\eta$ and polydispersity $\sigma$ of uniform-distribution polydisperse hard sphere . (a) Neural-network generator. (b) Percus-Yevick approximation. (c) Percus-Yevick approximation with $\beta$ correction.}
    \label{fig:gen_vs_PY_MSE_uni}
\end{figure}

Fig.~\ref{fig:gen_vs_PY_MSE_uni} shows a more systematic comparison between our neural-network model and the traditional method on the accuracy of scattering function generation. In which the size and color of the scatter indicated the mean square error $MSE=\left< (\log_{10}{I(Q)/I'(Q)})^2\right>_Q$ between the generated  scattering function $I'(Q)$ and the simulation calculated $I(Q)$. For our neural-network generator, the MSE is small across all different volume fraction and polydispersity. For the Percus-Yevick approximation, the error increases significantly with the increasing volume fraction and polydispersity. The $\beta$ correction reduces the error of the Percus-Yevick approximation but still significantly under performs compared to our neural-network model.

\subsection{Inference from scattering}

% fig 8 inference for pdType 1
We then turn to examine the inferrer part of our neural-network model that consist of the the encoder network that compress the entire scattering function curve $I(Q)$ into the latent variables, and the converter network that transform the latent variables to the predicted volume fraction $\eta'$ and polydispersity $\sigma'$.

\begin{figure}[!t]
    \centering
    \includegraphics[width=\linewidth]{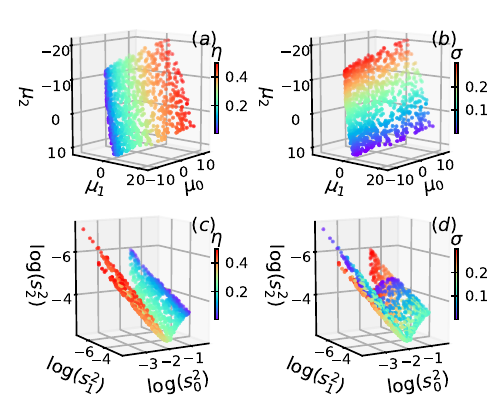}
    \caption{Distribution of volume fraction $\eta$ and polydispersity $\sigma$ in the latent space, for the testing set. (a) Volume fraction $\eta$ in the latent mean space $(\mu_0,\mu_1,\mu_2)$. (b) Polydispersity $\sigma$ in the latent mean space. (c) Volume fraction $\eta$ in the latent log variance space $(\log(s^2_0),\log(s^2_1),\log(s^2_2))$. (d) Polydispersity $\sigma$ in the latent log variance space.}
    \label{fig:parm_in_latent_space}
\end{figure}

First, we investigate the distribution of the system parameters including volume fraction and polydispersity in the latent space. As shown in Fig.~\ref{fig:parm_in_latent_space}, the system parameters are well-behaved in the latent space, such that their variation in the latent space are continuous and gradual, indicating the prevision mapping between the latent variables and the system parameters, especially highlighted by the distribution in the latent mean variable space $(\mu_0,\mu_1,\mu_2)$ as shown in Fig.~\ref{fig:parm_in_latent_space}(a) and (b). Meanwhile, the distribution in the latent log variance space is less important since the corresponding standard deviation $s$ are very small, indicating most of the information are encoded in the latent mean space.

\begin{figure}[!t]
    \centering
    \includegraphics[width=\linewidth]{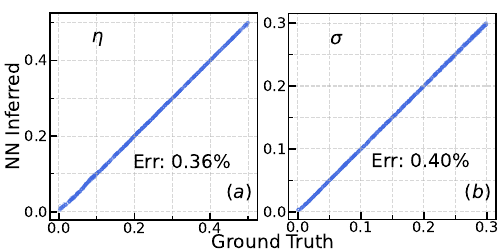}
    \caption{Comparison between the system parameters extracted from the scattering function using neural-network inferrer and the corresponding simulation calculated ground truth of uniform-distribution polydisperse hard spheres. (a) For volume fraction $\eta$. (b) For polydispersity $\sigma$.}
    \label{fig:inferrer_prediction_uni}
\end{figure}

Applying the inferrer that trained on the training set of $4000$ pairs of $I(Q)$ and $(\eta,\sigma)$ to the testing set, Fig.~\ref{fig:inferrer_prediction_uni} shows the comparison between the neural-network inferred system parameters and the ground truth input for the MD simulations. All of the scatter points are along the diagonal line and the relative error defined by $Err = |x-x'|/\left<x\right>$ for $x\in {\mu,\sigma}$ are very small. More importantly, unlike the traditional approach that requires a regression model for extracting these system parameters from the scattering function, our neural-network inferrer model can extract these information instantly since it is just once forward pass of the trained neural network.

\subsection{Other distributions}

\begin{figure}[!b]
    \centering
    \includegraphics[width=\linewidth]{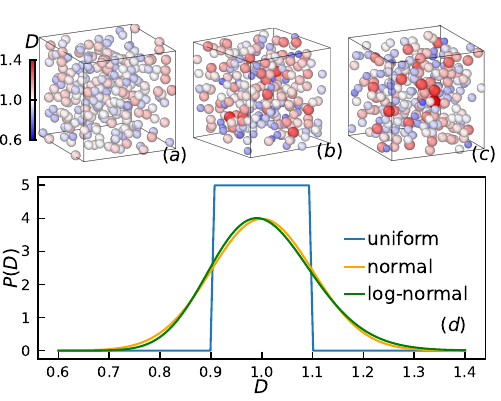}
    \caption{Illustration of the three different size distribution of 256 hard spheres with volume fraction $\eta=0.1$. (a) Uniform distribution in $D\sim U[D_0-\sigma,D_0+\sigma]$ with $D_0=1$ and $\sigma=0.1$. (b) Normal distribution $D\sim\mathcal{N}(D_0,\sigma)$ with mean $D_0=1$ and standard deviation $\sigma=0.1$. (c) Log-normal distribution $\log D \sim \mathcal{N}(\log{D_0},\sigma)$ with median $D_0=1 $ and $\sigma=0.1$. (d) Probability density function of three kinds of size distribution.}
    \label{fig:three_distribution}
\end{figure}
While the above detailed study are mainly focused on one type of particle size distribution, our approach is general, and can easily apply to any other polydispersity distribution. The commonly used ones are uniform, normal, and lognormal. Fig.~\ref{fig:three_distribution} shows sample configurations of the hard spheres and the diameter distribution of these three distribution. The diameter of the hard sphere particles are sampled from the corresponding distribution, parameterized by the polydispersity $\sigma$. For instance, the uniform distribution results in an uniform distribution of the particle diameter $D$ in the continuous interval $[D_0-\sigma, D_0 + \sigma]$ and we use $D_0=1$ as the natural unit. For the normal distribution, the diameter $D$ of each particle follow the normal distribution with mean $D_0=1$ and standard deviation $\sigma$, such that $D\sim\mathcal{N}(D_0,\sigma)$. Finally, for the lognormal distribution, the log of the particle diameter follows the normal distribution such that $\log D \sim \mathcal{N}(\log{D_0},\sigma)$, the $D_0=1$ become the median diameter.

Similar to the case of uniform distribution, we prepare dataset consist of 5,000 scattering function $I(Q)$ and system parameters: volume fraction $\eta$ and polydispersity $\sigma$ for each of the polydispersity type. We then repeat the training step for the neural network for each of these two new polydispersity distribution type. We examine both the generator and the inferrer network. 

\begin{figure}[!t]
    \centering
    \includegraphics[width=\linewidth]{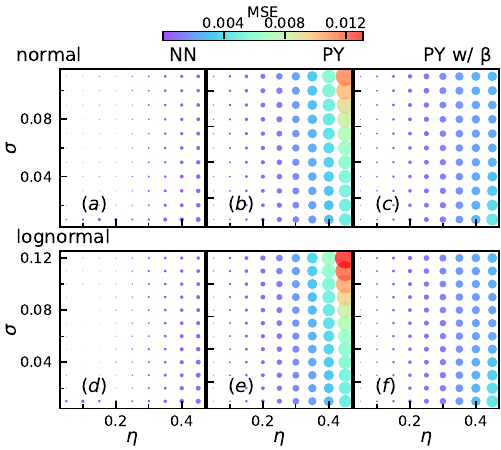}
    \caption{Mean square error $MSE=\left< (\log_{10}{I(Q)/I'(Q)})^2\right>_Q$ of the generated $I'(Q)$ using 3 different models for various volume fraction $\eta$ and polydispersity $\sigma$ of normal and lognormal-distribution polydisperse hard sphere. (a) Neural-network generator for normal distribution. (b) Percus-Yevick approximation for normal distribution. (c) Percus-Yevick approximation with $\beta$ correction for normal distribution. (d)-(f) Similar to (a)-(c) but for lognormal distribution.}
    \label{fig:gen_vs_PY_MSE_other}
\end{figure}

When it comes to generating the scattering functions, our model produces significant less discrepancy compared to the traditional methods, as shown in Fig.~\ref{fig:gen_vs_PY_MSE_other}, consisting with the results for uniform distribution.

\begin{figure}[!t]
    \centering
    \includegraphics[width=\linewidth]{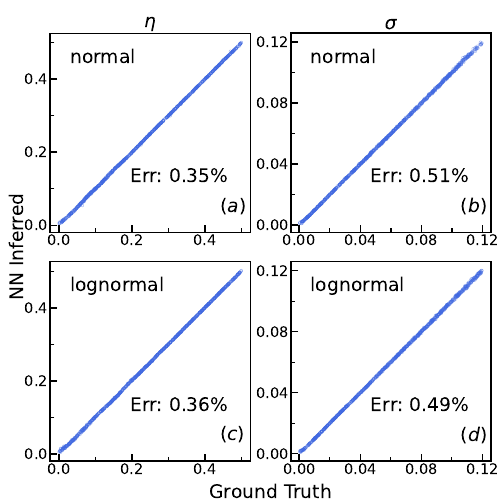}
    \caption{Comparison between the system parameters extracted from the scattering function using neural-network inferrer and the corresponding simulation calculated ground truth of normal and lognormal-distribution polydisperse hard spheres. (a) For volume fraction $\eta$ for normal distribution. (b) For polydispersity $\sigma$ for normal distribution. (c) and (d) Similar to (a) and (b) but for lognormal distribution.}
    \label{fig:inferrer_prediction_other}
\end{figure}

Furthermore, we also apply the inferrer network to extract volume fraction $\eta$ and polydispersity $\sigma$ from the scattering function for both the normal and lognormal distribution systems. As shown in Fig.~\ref{fig:inferrer_prediction_other}, the inferrer yields high-accuracy for parameter extraction.

\section{Summary}
In this work, we introduce a deep learning framework for not only generating scattering function of the polydisperse hard spheres, but also extracting system parameters including volume fraction and polydispersity directly from the scattering function. Our framework utilizes a neural network consisting of VAE for compressing the high dimensional scattering function vector $I(Q)$ into low dimensional latent variables, as well as two converter networks learning the mapping between the system parameters and the latent variables. We show that our generative model is significantly more accurate than the traditional methods and our inference model present high accuracy for parameter extraction from the scattering function.

% what results we have and significance (finding polydispersity from scattering)
We first generate and analyze the scattering data of the polydisperse hard spheres by carrying out MD simulation for different combination of volume fraction $\eta$ and polydispersity $\sigma$. Then, we build and train our neural network model using the scattering data paired with corresponding system parameters. From our trained neural network, we obtain a generator network that directly generate the scattering function from the system parameter, as well as a inferrer network that directly extract the system parameters from the scattering function. To benchmark our neural-network model, we compare our generator with the traditional Percus-Yevick approximation and its corresponding $\beta$ correction, our model yields low error cross all different volume fraction and polydispersity while the traditional methods suffer at high polydispersity and high volume fraction. And we compare the inferrer model extracted system parameters with the ground truth ones, it also has very high accuracy for a wide range of $\eta$ and $\sigma$. We apply these approach to three common polydispersity distribution including uniform, normal, and lognormal distribution. 

% future directions: (other colloidal system: polydisperse binary hard spheres, rods, where OZ really fails.
This work lays the groundwork for applying ML-based analysis to the scattering of polydisperse systems. Moving forward, \ld{our approach can be applied on experimental data, and both the generator and inferrer can be used under different conditions. When using the inferrer, one needs to firstly find the mean particle size for data normalization using dilute sample, then can just apply the inferrer for sample of higher concentration to achieve one-pass inference. When there is not dilute sample for reference or the instrument resolution is not high enough, one can combing the generator with a regression model to fit the experimental data\cite{tung2024inferring}. Our simulation also disgard the multiple scattering, which is recommended to be minimized in the real experimental setting by reducing the path length of the sample or by reducing the scattering contrast. Moreover,} this framework can be applied to even more intricate systems, such as charged colloids, binary colloids\cite{imhof1995experimental}, rods and charged rods\cite{schneider1987pair}, as well as other dense dispersion systems, where the strong coupling between the form factor and structure factor renders traditional analytic modeling ineffective. Since polydispersity is ubiquitous in nature, extending our framework to these cases will greatly improve the analytical power of the scattering technique.

\section*{Data Availability}
The code, data and trained model for this work are available at the GitHub repository \url{https://github.com/ljding94/Polydisperse_Sphere}

\begin{acknowledgments}
We thank Chi-Huan Tung, Wei-Ren Chen, and Jan-Michael Carrillo for fruitful discussion. This research used resources at the Center for Nanophase Materials Sciences, US Department of Energy (DOE) Office of Science User Facilities operated by Oak Ridge National Laboratory. This research was sponsored by the Laboratory Directed Research and Development Program of Oak Ridge National Laboratory, managed by UT-Battelle, LLC, for the US DOE. Computations used resources of the Oak Ridge Leadership Computing Facility, which is supported by the DOE Office of Science under contract No. DE-AC05-00OR22725. Application of machine learning to soft matter was supported by the US DOE, Office of Science, Office of Basic Energy Sciences Data, Artificial Intelligence and Machine Learning at DOE Scientific User Facilities Program under award No. 34532.
\end{acknowledgments}

%\nocite{*}

\bibliography{reference}% Produces the bibliography via BibTeX.

%aipnum4-2.bst 2019-01-14 (MD) hand-edited version of apsrev4-1.bst
\begin{thebibliography}{51}%
\makeatletter
\providecommand \@ifxundefined [1]{%
 \@ifx{#1\undefined}
}%
\providecommand \@ifnum [1]{%
 \ifnum #1\expandafter \@firstoftwo
 \else \expandafter \@secondoftwo
 \fi
}%
\providecommand \@ifx [1]{%
 \ifx #1\expandafter \@firstoftwo
 \else \expandafter \@secondoftwo
 \fi
}%
\providecommand \natexlab [1]{#1}%
\providecommand \enquote  [1]{``#1''}%
\providecommand \bibnamefont  [1]{#1}%
\providecommand \bibfnamefont [1]{#1}%
\providecommand \citenamefont [1]{#1}%
\providecommand \href@noop [0]{\@secondoftwo}%
\providecommand \href [0]{\begingroup \@sanitize@url \@href}%
\providecommand \@href[1]{\@@startlink{#1}\@@href}%
\providecommand \@@href[1]{\endgroup#1\@@endlink}%
\providecommand \@sanitize@url [0]{\catcode `\\12\catcode `\$12\catcode
  `\&12\catcode `\#12\catcode `\^12\catcode `\_12\catcode `\%12\relax}%
\providecommand \@@startlink[1]{}%
\providecommand \@@endlink[0]{}%
\providecommand \url  [0]{\begingroup\@sanitize@url \@url }%
\providecommand \@url [1]{\endgroup\@href {#1}{\urlprefix }}%
\providecommand \urlprefix  [0]{URL }%
\providecommand \Eprint [0]{\href }%
\providecommand \doibase [0]{https://doi.org/}%
\providecommand \selectlanguage [0]{\@gobble}%
\providecommand \bibinfo  [0]{\@secondoftwo}%
\providecommand \bibfield  [0]{\@secondoftwo}%
\providecommand \translation [1]{[#1]}%
\providecommand \BibitemOpen [0]{}%
\providecommand \bibitemStop [0]{}%
\providecommand \bibitemNoStop [0]{.\EOS\space}%
\providecommand \EOS [0]{\spacefactor3000\relax}%
\providecommand \BibitemShut  [1]{\csname bibitem#1\endcsname}%
\let\auto@bib@innerbib\@empty
%</preamble>
\bibitem [{\citenamefont {Russel}\ \emph {et~al.}(1991)\citenamefont {Russel},
  \citenamefont {Russel}, \citenamefont {Saville},\ and\ \citenamefont
  {Schowalter}}]{russel1991colloidal}%
  \BibitemOpen
  \bibfield  {author} {\bibinfo {author} {\bibfnamefont {W.~B.}\ \bibnamefont
  {Russel}}, \bibinfo {author} {\bibfnamefont {W.}~\bibnamefont {Russel}},
  \bibinfo {author} {\bibfnamefont {D.~A.}\ \bibnamefont {Saville}},\ and\
  \bibinfo {author} {\bibfnamefont {W.~R.}\ \bibnamefont {Schowalter}},\
  }\href@noop {} {\emph {\bibinfo {title} {Colloidal dispersions}}}\ (\bibinfo
  {publisher} {Cambridge university press},\ \bibinfo {year}
  {1991})\BibitemShut {NoStop}%
\bibitem [{\citenamefont {Lindner}\ and\ \citenamefont
  {Zemb}(2002)}]{lindner2002neutrons}%
  \BibitemOpen
  \bibfield  {author} {\bibinfo {author} {\bibfnamefont {P.}~\bibnamefont
  {Lindner}}\ and\ \bibinfo {author} {\bibfnamefont {T.}~\bibnamefont {Zemb}},\
  }\href@noop {} {\emph {\bibinfo {title} {Neutrons, X-rays and Light:
  Scattering Methods Applied to Soft Condensed Matter}}},\ North-Holland Delta
  Series\ (\bibinfo  {publisher} {Elsevier Science},\ \bibinfo {year}
  {2002})\BibitemShut {NoStop}%
\bibitem [{\citenamefont {Shibayama}(2011)}]{shibayama2011small}%
  \BibitemOpen
  \bibfield  {author} {\bibinfo {author} {\bibfnamefont {M.}~\bibnamefont
  {Shibayama}},\ }\bibfield  {title} {\enquote {\bibinfo {title} {Small-angle
  neutron scattering on polymer gels: phase behavior, inhomogeneities and
  deformation mechanisms},}\ }\href@noop {} {\bibfield  {journal} {\bibinfo
  {journal} {Polymer journal}\ }\textbf {\bibinfo {volume} {43}},\ \bibinfo
  {pages} {18--34} (\bibinfo {year} {2011})}\BibitemShut {NoStop}%
\bibitem [{\citenamefont {Chen}(1986)}]{chen1986small}%
  \BibitemOpen
  \bibfield  {author} {\bibinfo {author} {\bibfnamefont {S.-H.}\ \bibnamefont
  {Chen}},\ }\bibfield  {title} {\enquote {\bibinfo {title} {Small angle
  neutron scattering studies of the structure and interaction in micellar and
  microemulsion systems},}\ }\href@noop {} {\bibfield  {journal} {\bibinfo
  {journal} {Annual Review of Physical Chemistry}\ }\textbf {\bibinfo {volume}
  {37}},\ \bibinfo {pages} {351--399} (\bibinfo {year} {1986})}\BibitemShut
  {NoStop}%
\bibitem [{\citenamefont {Chu}\ and\ \citenamefont
  {Hsiao}(2001)}]{chu2001small}%
  \BibitemOpen
  \bibfield  {author} {\bibinfo {author} {\bibfnamefont {B.}~\bibnamefont
  {Chu}}\ and\ \bibinfo {author} {\bibfnamefont {B.~S.}\ \bibnamefont
  {Hsiao}},\ }\bibfield  {title} {\enquote {\bibinfo {title} {Small-angle x-ray
  scattering of polymers},}\ }\href@noop {} {\bibfield  {journal} {\bibinfo
  {journal} {Chemical reviews}\ }\textbf {\bibinfo {volume} {101}},\ \bibinfo
  {pages} {1727--1762} (\bibinfo {year} {2001})}\BibitemShut {NoStop}%
\bibitem [{\citenamefont {Debye}(1947)}]{debye1947molecular}%
  \BibitemOpen
  \bibfield  {author} {\bibinfo {author} {\bibfnamefont {P.}~\bibnamefont
  {Debye}},\ }\bibfield  {title} {\enquote {\bibinfo {title} {Molecular-weight
  determination by light scattering.}}\ }\href@noop {} {\bibfield  {journal}
  {\bibinfo  {journal} {The Journal of Physical Chemistry}\ }\textbf {\bibinfo
  {volume} {51}},\ \bibinfo {pages} {18--32} (\bibinfo {year}
  {1947})}\BibitemShut {NoStop}%
\bibitem [{\citenamefont {Sch{\"a}rtl}(2007)}]{schartl2007light}%
  \BibitemOpen
  \bibfield  {author} {\bibinfo {author} {\bibfnamefont {W.}~\bibnamefont
  {Sch{\"a}rtl}},\ }\href@noop {} {\emph {\bibinfo {title} {Light scattering
  from polymer solutions and nanoparticle dispersions}}}\ (\bibinfo
  {publisher} {Springer Science \& Business Media},\ \bibinfo {year}
  {2007})\BibitemShut {NoStop}%
\bibitem [{\citenamefont {Mittelbach}\ and\ \citenamefont
  {Glatter}(1998)}]{mittelbach1998direct}%
  \BibitemOpen
  \bibfield  {author} {\bibinfo {author} {\bibfnamefont {R.}~\bibnamefont
  {Mittelbach}}\ and\ \bibinfo {author} {\bibfnamefont {O.}~\bibnamefont
  {Glatter}},\ }\bibfield  {title} {\enquote {\bibinfo {title} {Direct
  structure analysis of small-angle scattering data from polydisperse colloidal
  particles},}\ }\href@noop {} {\bibfield  {journal} {\bibinfo  {journal}
  {Applied Crystallography}\ }\textbf {\bibinfo {volume} {31}},\ \bibinfo
  {pages} {600--608} (\bibinfo {year} {1998})}\BibitemShut {NoStop}%
\bibitem [{\citenamefont {Huang}\ \emph {et~al.}(2023)\citenamefont {Huang},
  \citenamefont {Tung}, \citenamefont {Porcar}, \citenamefont {Wang},
  \citenamefont {Shinohara}, \citenamefont {Do},\ and\ \citenamefont
  {Chen}}]{huang2023model}%
  \BibitemOpen
  \bibfield  {author} {\bibinfo {author} {\bibfnamefont {G.-R.}\ \bibnamefont
  {Huang}}, \bibinfo {author} {\bibfnamefont {C.-H.}\ \bibnamefont {Tung}},
  \bibinfo {author} {\bibfnamefont {L.}~\bibnamefont {Porcar}}, \bibinfo
  {author} {\bibfnamefont {Y.}~\bibnamefont {Wang}}, \bibinfo {author}
  {\bibfnamefont {Y.}~\bibnamefont {Shinohara}}, \bibinfo {author}
  {\bibfnamefont {C.}~\bibnamefont {Do}},\ and\ \bibinfo {author}
  {\bibfnamefont {W.-R.}\ \bibnamefont {Chen}},\ }\bibfield  {title} {\enquote
  {\bibinfo {title} {Model-free approach for profiling of polydisperse soft
  matter using small angle scattering},}\ }\href@noop {} {\bibfield  {journal}
  {\bibinfo  {journal} {Macromolecules}\ }\textbf {\bibinfo {volume} {56}},\
  \bibinfo {pages} {6436--6443} (\bibinfo {year} {2023})}\BibitemShut {NoStop}%
\bibitem [{\citenamefont {Weyerich}, \citenamefont {Brunner-Popela},\ and\
  \citenamefont {Glatter}(1999)}]{weyerich1999small}%
  \BibitemOpen
  \bibfield  {author} {\bibinfo {author} {\bibfnamefont {B.}~\bibnamefont
  {Weyerich}}, \bibinfo {author} {\bibfnamefont {J.}~\bibnamefont
  {Brunner-Popela}},\ and\ \bibinfo {author} {\bibfnamefont {O.}~\bibnamefont
  {Glatter}},\ }\bibfield  {title} {\enquote {\bibinfo {title} {Small-angle
  scattering of interacting particles. ii. generalized indirect fourier
  transformation under consideration of the effective structure factor for
  polydisperse systems},}\ }\href@noop {} {\bibfield  {journal} {\bibinfo
  {journal} {Applied Crystallography}\ }\textbf {\bibinfo {volume} {32}},\
  \bibinfo {pages} {197--209} (\bibinfo {year} {1999})}\BibitemShut {NoStop}%
\bibitem [{\citenamefont {Salgi}\ and\ \citenamefont
  {Rajagopalan}(1993)}]{salgi1993polydispersity}%
  \BibitemOpen
  \bibfield  {author} {\bibinfo {author} {\bibfnamefont {P.}~\bibnamefont
  {Salgi}}\ and\ \bibinfo {author} {\bibfnamefont {R.}~\bibnamefont
  {Rajagopalan}},\ }\bibfield  {title} {\enquote {\bibinfo {title}
  {Polydispersity in colloids: implications to static structure and
  scattering},}\ }\href@noop {} {\bibfield  {journal} {\bibinfo  {journal}
  {Advances in Colloid and Interface science}\ }\textbf {\bibinfo {volume}
  {43}},\ \bibinfo {pages} {169--288} (\bibinfo {year} {1993})}\BibitemShut
  {NoStop}%
\bibitem [{\citenamefont {Wagner}\ and\ \citenamefont
  {Woutersen}(1994)}]{wagner1994viscosity}%
  \BibitemOpen
  \bibfield  {author} {\bibinfo {author} {\bibfnamefont {N.}~\bibnamefont
  {Wagner}}\ and\ \bibinfo {author} {\bibfnamefont {A.~M.}\ \bibnamefont
  {Woutersen}},\ }\bibfield  {title} {\enquote {\bibinfo {title} {The viscosity
  of bimodal and polydisperse suspensions of hard spheres in the dilute
  limit},}\ }\href@noop {} {\bibfield  {journal} {\bibinfo  {journal} {Journal
  of Fluid Mechanics}\ }\textbf {\bibinfo {volume} {278}},\ \bibinfo {pages}
  {267--287} (\bibinfo {year} {1994})}\BibitemShut {NoStop}%
\bibitem [{\citenamefont {Eckert}, \citenamefont {Schmidt},\ and\ \citenamefont
  {de~Las~Heras}(2022)}]{eckert2022sedimentation}%
  \BibitemOpen
  \bibfield  {author} {\bibinfo {author} {\bibfnamefont {T.}~\bibnamefont
  {Eckert}}, \bibinfo {author} {\bibfnamefont {M.}~\bibnamefont {Schmidt}},\
  and\ \bibinfo {author} {\bibfnamefont {D.}~\bibnamefont {de~Las~Heras}},\
  }\bibfield  {title} {\enquote {\bibinfo {title} {Sedimentation path theory
  for mass-polydisperse colloidal systems},}\ }\href@noop {} {\bibfield
  {journal} {\bibinfo  {journal} {The Journal of Chemical Physics}\ }\textbf
  {\bibinfo {volume} {157}} (\bibinfo {year} {2022})}\BibitemShut {NoStop}%
\bibitem [{\citenamefont {Pusey}\ and\ \citenamefont
  {Van~Megen}(1986)}]{pusey1986phase}%
  \BibitemOpen
  \bibfield  {author} {\bibinfo {author} {\bibfnamefont {P.~N.}\ \bibnamefont
  {Pusey}}\ and\ \bibinfo {author} {\bibfnamefont {W.}~\bibnamefont
  {Van~Megen}},\ }\bibfield  {title} {\enquote {\bibinfo {title} {Phase
  behaviour of concentrated suspensions of nearly hard colloidal spheres},}\
  }\href@noop {} {\bibfield  {journal} {\bibinfo  {journal} {Nature}\ }\textbf
  {\bibinfo {volume} {320}},\ \bibinfo {pages} {340--342} (\bibinfo {year}
  {1986})}\BibitemShut {NoStop}%
\bibitem [{\citenamefont {Percus}\ and\ \citenamefont
  {Yevick}(1958)}]{percus1958analysis}%
  \BibitemOpen
  \bibfield  {author} {\bibinfo {author} {\bibfnamefont {J.~K.}\ \bibnamefont
  {Percus}}\ and\ \bibinfo {author} {\bibfnamefont {G.~J.}\ \bibnamefont
  {Yevick}},\ }\bibfield  {title} {\enquote {\bibinfo {title} {Analysis of
  classical statistical mechanics by means of collective coordinates},}\
  }\href@noop {} {\bibfield  {journal} {\bibinfo  {journal} {Physical Review}\
  }\textbf {\bibinfo {volume} {110}},\ \bibinfo {pages} {1} (\bibinfo {year}
  {1958})}\BibitemShut {NoStop}%
\bibitem [{\citenamefont {Wertheim}(1963)}]{wertheim1963exact}%
  \BibitemOpen
  \bibfield  {author} {\bibinfo {author} {\bibfnamefont {M.}~\bibnamefont
  {Wertheim}},\ }\bibfield  {title} {\enquote {\bibinfo {title} {Exact solution
  of the percus-yevick integral equation for hard spheres},}\ }\href@noop {}
  {\bibfield  {journal} {\bibinfo  {journal} {Physical Review Letters}\
  }\textbf {\bibinfo {volume} {10}},\ \bibinfo {pages} {321} (\bibinfo {year}
  {1963})}\BibitemShut {NoStop}%
\bibitem [{\citenamefont {Katzav}, \citenamefont {Berdichevsky},\ and\
  \citenamefont {Schwartz}(2019)}]{katzav2019random}%
  \BibitemOpen
  \bibfield  {author} {\bibinfo {author} {\bibfnamefont {E.}~\bibnamefont
  {Katzav}}, \bibinfo {author} {\bibfnamefont {R.}~\bibnamefont
  {Berdichevsky}},\ and\ \bibinfo {author} {\bibfnamefont {M.}~\bibnamefont
  {Schwartz}},\ }\bibfield  {title} {\enquote {\bibinfo {title} {Random close
  packing from hard-sphere percus-yevick theory},}\ }\href@noop {} {\bibfield
  {journal} {\bibinfo  {journal} {Physical Review E}\ }\textbf {\bibinfo
  {volume} {99}},\ \bibinfo {pages} {012146} (\bibinfo {year}
  {2019})}\BibitemShut {NoStop}%
\bibitem [{\citenamefont {Zaccone}(2022)}]{zaccone2022explicit}%
  \BibitemOpen
  \bibfield  {author} {\bibinfo {author} {\bibfnamefont {A.}~\bibnamefont
  {Zaccone}},\ }\bibfield  {title} {\enquote {\bibinfo {title} {Explicit
  analytical solution for random close packing in d= 2 and d= 3},}\ }\href@noop
  {} {\bibfield  {journal} {\bibinfo  {journal} {Physical Review Letters}\
  }\textbf {\bibinfo {volume} {128}},\ \bibinfo {pages} {028002} (\bibinfo
  {year} {2022})}\BibitemShut {NoStop}%
\bibitem [{\citenamefont {Anzivino}\ \emph {et~al.}(2023)\citenamefont
  {Anzivino}, \citenamefont {Casiulis}, \citenamefont {Zhang}, \citenamefont
  {Moussa}, \citenamefont {Martiniani},\ and\ \citenamefont
  {Zaccone}}]{anzivino2023estimating}%
  \BibitemOpen
  \bibfield  {author} {\bibinfo {author} {\bibfnamefont {C.}~\bibnamefont
  {Anzivino}}, \bibinfo {author} {\bibfnamefont {M.}~\bibnamefont {Casiulis}},
  \bibinfo {author} {\bibfnamefont {T.}~\bibnamefont {Zhang}}, \bibinfo
  {author} {\bibfnamefont {A.~S.}\ \bibnamefont {Moussa}}, \bibinfo {author}
  {\bibfnamefont {S.}~\bibnamefont {Martiniani}},\ and\ \bibinfo {author}
  {\bibfnamefont {A.}~\bibnamefont {Zaccone}},\ }\bibfield  {title} {\enquote
  {\bibinfo {title} {Estimating random close packing in polydisperse and
  bidisperse hard spheres via an equilibrium model of crowding},}\ }\href@noop
  {} {\bibfield  {journal} {\bibinfo  {journal} {The Journal of Chemical
  Physics}\ }\textbf {\bibinfo {volume} {158}} (\bibinfo {year}
  {2023})}\BibitemShut {NoStop}%
\bibitem [{\citenamefont {Murphy}(2012)}]{murphy2012machine}%
  \BibitemOpen
  \bibfield  {author} {\bibinfo {author} {\bibfnamefont {K.~P.}\ \bibnamefont
  {Murphy}},\ }\href@noop {} {\emph {\bibinfo {title} {Machine learning: a
  probabilistic perspective}}}\ (\bibinfo  {publisher} {MIT press},\ \bibinfo
  {year} {2012})\BibitemShut {NoStop}%
\bibitem [{\citenamefont {Carleo}\ \emph {et~al.}(2019)\citenamefont {Carleo},
  \citenamefont {Cirac}, \citenamefont {Cranmer}, \citenamefont {Daudet},
  \citenamefont {Schuld}, \citenamefont {Tishby}, \citenamefont
  {Vogt-Maranto},\ and\ \citenamefont {Zdeborov{\'a}}}]{carleo2019machine}%
  \BibitemOpen
  \bibfield  {author} {\bibinfo {author} {\bibfnamefont {G.}~\bibnamefont
  {Carleo}}, \bibinfo {author} {\bibfnamefont {I.}~\bibnamefont {Cirac}},
  \bibinfo {author} {\bibfnamefont {K.}~\bibnamefont {Cranmer}}, \bibinfo
  {author} {\bibfnamefont {L.}~\bibnamefont {Daudet}}, \bibinfo {author}
  {\bibfnamefont {M.}~\bibnamefont {Schuld}}, \bibinfo {author} {\bibfnamefont
  {N.}~\bibnamefont {Tishby}}, \bibinfo {author} {\bibfnamefont
  {L.}~\bibnamefont {Vogt-Maranto}},\ and\ \bibinfo {author} {\bibfnamefont
  {L.}~\bibnamefont {Zdeborov{\'a}}},\ }\bibfield  {title} {\enquote {\bibinfo
  {title} {Machine learning and the physical sciences},}\ }\href@noop {}
  {\bibfield  {journal} {\bibinfo  {journal} {Reviews of Modern Physics}\
  }\textbf {\bibinfo {volume} {91}},\ \bibinfo {pages} {045002} (\bibinfo
  {year} {2019})}\BibitemShut {NoStop}%
\bibitem [{\citenamefont {Goodfellow}\ \emph {et~al.}(2016)\citenamefont
  {Goodfellow}, \citenamefont {Bengio}, \citenamefont {Courville},\ and\
  \citenamefont {Bengio}}]{goodfellow2016deep}%
  \BibitemOpen
  \bibfield  {author} {\bibinfo {author} {\bibfnamefont {I.}~\bibnamefont
  {Goodfellow}}, \bibinfo {author} {\bibfnamefont {Y.}~\bibnamefont {Bengio}},
  \bibinfo {author} {\bibfnamefont {A.}~\bibnamefont {Courville}},\ and\
  \bibinfo {author} {\bibfnamefont {Y.}~\bibnamefont {Bengio}},\ }\href@noop {}
  {\emph {\bibinfo {title} {Deep learning}}},\ Vol.~\bibinfo {volume} {1}\
  (\bibinfo  {publisher} {MIT press Cambridge},\ \bibinfo {year}
  {2016})\BibitemShut {NoStop}%
\bibitem [{\citenamefont {LeCun}, \citenamefont {Bengio},\ and\ \citenamefont
  {Hinton}(2015)}]{lecun2015deep}%
  \BibitemOpen
  \bibfield  {author} {\bibinfo {author} {\bibfnamefont {Y.}~\bibnamefont
  {LeCun}}, \bibinfo {author} {\bibfnamefont {Y.}~\bibnamefont {Bengio}},\ and\
  \bibinfo {author} {\bibfnamefont {G.}~\bibnamefont {Hinton}},\ }\bibfield
  {title} {\enquote {\bibinfo {title} {Deep learning},}\ }\href@noop {}
  {\bibfield  {journal} {\bibinfo  {journal} {nature}\ }\textbf {\bibinfo
  {volume} {521}},\ \bibinfo {pages} {436--444} (\bibinfo {year}
  {2015})}\BibitemShut {NoStop}%
\bibitem [{\citenamefont {Chang}\ \emph {et~al.}(2022)\citenamefont {Chang},
  \citenamefont {Tung}, \citenamefont {Chang}, \citenamefont {Carrillo},
  \citenamefont {Wang}, \citenamefont {Sumpter}, \citenamefont {Huang},
  \citenamefont {Do},\ and\ \citenamefont {Chen}}]{chang2022machine}%
  \BibitemOpen
  \bibfield  {author} {\bibinfo {author} {\bibfnamefont {M.-C.}\ \bibnamefont
  {Chang}}, \bibinfo {author} {\bibfnamefont {C.-H.}\ \bibnamefont {Tung}},
  \bibinfo {author} {\bibfnamefont {S.-Y.}\ \bibnamefont {Chang}}, \bibinfo
  {author} {\bibfnamefont {J.~M.}\ \bibnamefont {Carrillo}}, \bibinfo {author}
  {\bibfnamefont {Y.}~\bibnamefont {Wang}}, \bibinfo {author} {\bibfnamefont
  {B.~G.}\ \bibnamefont {Sumpter}}, \bibinfo {author} {\bibfnamefont {G.-R.}\
  \bibnamefont {Huang}}, \bibinfo {author} {\bibfnamefont {C.}~\bibnamefont
  {Do}},\ and\ \bibinfo {author} {\bibfnamefont {W.-R.}\ \bibnamefont {Chen}},\
  }\bibfield  {title} {\enquote {\bibinfo {title} {A machine learning inversion
  scheme for determining interaction from scattering},}\ }\href@noop {}
  {\bibfield  {journal} {\bibinfo  {journal} {Communications Physics}\ }\textbf
  {\bibinfo {volume} {5}},\ \bibinfo {pages} {46} (\bibinfo {year}
  {2022})}\BibitemShut {NoStop}%
\bibitem [{\citenamefont {Tung}\ \emph {et~al.}(2024)\citenamefont {Tung},
  \citenamefont {Chen}, \citenamefont {Chen}, \citenamefont {Huang},
  \citenamefont {Porcar}, \citenamefont {Chang}, \citenamefont {Carrillo},
  \citenamefont {Wang}, \citenamefont {Sumpter}, \citenamefont {Shinohara}
  \emph {et~al.}}]{tung2024inferring}%
  \BibitemOpen
  \bibfield  {author} {\bibinfo {author} {\bibfnamefont {C.-H.}\ \bibnamefont
  {Tung}}, \bibinfo {author} {\bibfnamefont {M.-Z.}\ \bibnamefont {Chen}},
  \bibinfo {author} {\bibfnamefont {H.-L.}\ \bibnamefont {Chen}}, \bibinfo
  {author} {\bibfnamefont {G.-R.}\ \bibnamefont {Huang}}, \bibinfo {author}
  {\bibfnamefont {L.}~\bibnamefont {Porcar}}, \bibinfo {author} {\bibfnamefont
  {M.-C.}\ \bibnamefont {Chang}}, \bibinfo {author} {\bibfnamefont {J.-M.}\
  \bibnamefont {Carrillo}}, \bibinfo {author} {\bibfnamefont {Y.}~\bibnamefont
  {Wang}}, \bibinfo {author} {\bibfnamefont {B.~G.}\ \bibnamefont {Sumpter}},
  \bibinfo {author} {\bibfnamefont {Y.}~\bibnamefont {Shinohara}}, \emph
  {et~al.},\ }\bibfield  {title} {\enquote {\bibinfo {title} {Inferring
  effective electrostatic interaction of charge-stabilized colloids from
  scattering using deep learning},}\ }\href@noop {} {\bibfield  {journal}
  {\bibinfo  {journal} {Applied Crystallography}\ }\textbf {\bibinfo {volume}
  {57}} (\bibinfo {year} {2024})}\BibitemShut {NoStop}%
\bibitem [{\citenamefont {Ding}, \citenamefont {Chen},\ and\ \citenamefont
  {Do}(2025)}]{ding2025colloids}%
  \BibitemOpen
  \bibfield  {author} {\bibinfo {author} {\bibfnamefont {L.}~\bibnamefont
  {Ding}}, \bibinfo {author} {\bibfnamefont {Y.}~\bibnamefont {Chen}},\ and\
  \bibinfo {author} {\bibfnamefont {C.}~\bibnamefont {Do}},\ }\bibfield
  {title} {\enquote {\bibinfo {title} {Machine-learning-informed scattering
  correlation analysis of sheared colloids},}\ }\href@noop {} {\bibfield
  {journal} {\bibinfo  {journal} {Applied Crystallography}\ }\textbf {\bibinfo
  {volume} {58}},\ \bibinfo {pages} {992--999} (\bibinfo {year}
  {2025})}\BibitemShut {NoStop}%
\bibitem [{\citenamefont {Tung}\ \emph
  {et~al.}(2025{\natexlab{a}})\citenamefont {Tung}, \citenamefont {Ding},
  \citenamefont {Chang}, \citenamefont {Huang}, \citenamefont {Porcar},
  \citenamefont {Wang}, \citenamefont {Carrillo}, \citenamefont {Sumpter},
  \citenamefont {Shinohara}, \citenamefont {Do} \emph
  {et~al.}}]{tung2025scattering}%
  \BibitemOpen
  \bibfield  {author} {\bibinfo {author} {\bibfnamefont {C.-H.}\ \bibnamefont
  {Tung}}, \bibinfo {author} {\bibfnamefont {L.}~\bibnamefont {Ding}}, \bibinfo
  {author} {\bibfnamefont {M.-C.}\ \bibnamefont {Chang}}, \bibinfo {author}
  {\bibfnamefont {G.-R.}\ \bibnamefont {Huang}}, \bibinfo {author}
  {\bibfnamefont {L.}~\bibnamefont {Porcar}}, \bibinfo {author} {\bibfnamefont
  {Y.}~\bibnamefont {Wang}}, \bibinfo {author} {\bibfnamefont {J.-M.~Y.}\
  \bibnamefont {Carrillo}}, \bibinfo {author} {\bibfnamefont {B.~G.}\
  \bibnamefont {Sumpter}}, \bibinfo {author} {\bibfnamefont {Y.}~\bibnamefont
  {Shinohara}}, \bibinfo {author} {\bibfnamefont {C.}~\bibnamefont {Do}}, \emph
  {et~al.},\ }\bibfield  {title} {\enquote {\bibinfo {title} {Scattering-based
  structural inversion of soft materials via kolmogorov--arnold networks},}\
  }\href@noop {} {\bibfield  {journal} {\bibinfo  {journal} {The Journal of
  Chemical Physics}\ }\textbf {\bibinfo {volume} {162}} (\bibinfo {year}
  {2025}{\natexlab{a}})}\BibitemShut {NoStop}%
\bibitem [{\citenamefont {Tung}\ \emph
  {et~al.}(2025{\natexlab{b}})\citenamefont {Tung}, \citenamefont {Ding},
  \citenamefont {Huang}, \citenamefont {Porcar}, \citenamefont {Shinohara},
  \citenamefont {Sumpter}, \citenamefont {Do},\ and\ \citenamefont
  {Chen}}]{tung2025insights}%
  \BibitemOpen
  \bibfield  {author} {\bibinfo {author} {\bibfnamefont {C.-H.}\ \bibnamefont
  {Tung}}, \bibinfo {author} {\bibfnamefont {L.}~\bibnamefont {Ding}}, \bibinfo
  {author} {\bibfnamefont {G.-R.}\ \bibnamefont {Huang}}, \bibinfo {author}
  {\bibfnamefont {L.}~\bibnamefont {Porcar}}, \bibinfo {author} {\bibfnamefont
  {Y.}~\bibnamefont {Shinohara}}, \bibinfo {author} {\bibfnamefont {B.~G.}\
  \bibnamefont {Sumpter}}, \bibinfo {author} {\bibfnamefont {C.}~\bibnamefont
  {Do}},\ and\ \bibinfo {author} {\bibfnamefont {W.-R.}\ \bibnamefont {Chen}},\
  }\bibfield  {title} {\enquote {\bibinfo {title} {Insights into distorted
  lamellar phases with small-angle scattering and machine learning},}\
  }\href@noop {} {\bibfield  {journal} {\bibinfo  {journal} {Applied
  Crystallography}\ }\textbf {\bibinfo {volume} {58}} (\bibinfo {year}
  {2025}{\natexlab{b}})}\BibitemShut {NoStop}%
\bibitem [{\citenamefont {Ding}\ \emph {et~al.}(2024)\citenamefont {Ding},
  \citenamefont {Tung}, \citenamefont {Sumpter}, \citenamefont {Chen},\ and\
  \citenamefont {Do}}]{ding2024mechanical}%
  \BibitemOpen
  \bibfield  {author} {\bibinfo {author} {\bibfnamefont {L.}~\bibnamefont
  {Ding}}, \bibinfo {author} {\bibfnamefont {C.-H.}\ \bibnamefont {Tung}},
  \bibinfo {author} {\bibfnamefont {B.~G.}\ \bibnamefont {Sumpter}}, \bibinfo
  {author} {\bibfnamefont {W.-R.}\ \bibnamefont {Chen}},\ and\ \bibinfo
  {author} {\bibfnamefont {C.}~\bibnamefont {Do}},\ }\bibfield  {title}
  {\enquote {\bibinfo {title} {Machine learning inversion from scattering for
  mechanically driven polymers},}\ }\href@noop {} {\bibfield  {journal}
  {\bibinfo  {journal} {arXiv preprint arXiv:2410.05574}\ } (\bibinfo {year}
  {2024})}\BibitemShut {NoStop}%
\bibitem [{\citenamefont {Ding}\ \emph
  {et~al.}(2025{\natexlab{a}})\citenamefont {Ding}, \citenamefont {Tung},
  \citenamefont {Carrillo}, \citenamefont {Chen},\ and\ \citenamefont
  {Do}}]{ding2025charge}%
  \BibitemOpen
  \bibfield  {author} {\bibinfo {author} {\bibfnamefont {L.}~\bibnamefont
  {Ding}}, \bibinfo {author} {\bibfnamefont {C.-H.}\ \bibnamefont {Tung}},
  \bibinfo {author} {\bibfnamefont {J.~M.~Y.}\ \bibnamefont {Carrillo}},
  \bibinfo {author} {\bibfnamefont {W.-R.}\ \bibnamefont {Chen}},\ and\
  \bibinfo {author} {\bibfnamefont {C.}~\bibnamefont {Do}},\ }\bibfield
  {title} {\enquote {\bibinfo {title} {Machine learning inversion from
  small-angle scattering for charged polymers},}\ }\href@noop {} {\bibfield
  {journal} {\bibinfo  {journal} {Digital Discovery}\ } (\bibinfo {year}
  {2025}{\natexlab{a}})}\BibitemShut {NoStop}%
\bibitem [{\citenamefont {Ding}\ \emph
  {et~al.}(2025{\natexlab{b}})\citenamefont {Ding}, \citenamefont {Tung},
  \citenamefont {Cao}, \citenamefont {Ye}, \citenamefont {Gu}, \citenamefont
  {Xia}, \citenamefont {Chen},\ and\ \citenamefont {Do}}]{ding2025ladder}%
  \BibitemOpen
  \bibfield  {author} {\bibinfo {author} {\bibfnamefont {L.}~\bibnamefont
  {Ding}}, \bibinfo {author} {\bibfnamefont {C.-H.}\ \bibnamefont {Tung}},
  \bibinfo {author} {\bibfnamefont {Z.}~\bibnamefont {Cao}}, \bibinfo {author}
  {\bibfnamefont {Z.}~\bibnamefont {Ye}}, \bibinfo {author} {\bibfnamefont
  {X.}~\bibnamefont {Gu}}, \bibinfo {author} {\bibfnamefont {Y.}~\bibnamefont
  {Xia}}, \bibinfo {author} {\bibfnamefont {W.-R.}\ \bibnamefont {Chen}},\ and\
  \bibinfo {author} {\bibfnamefont {C.}~\bibnamefont {Do}},\ }\bibfield
  {title} {\enquote {\bibinfo {title} {Machine learning-assisted profiling of a
  kinked ladder polymer structure using scattering},}\ }\href@noop {}
  {\bibfield  {journal} {\bibinfo  {journal} {Digital Discovery}\ } (\bibinfo
  {year} {2025}{\natexlab{b}})}\BibitemShut {NoStop}%
\bibitem [{\citenamefont {Ding}\ \emph
  {et~al.}(2025{\natexlab{c}})\citenamefont {Ding}, \citenamefont {Tung},
  \citenamefont {Sumpter}, \citenamefont {Chen},\ and\ \citenamefont
  {Do}}]{ding2025deciphering}%
  \BibitemOpen
  \bibfield  {author} {\bibinfo {author} {\bibfnamefont {L.}~\bibnamefont
  {Ding}}, \bibinfo {author} {\bibfnamefont {C.-H.}\ \bibnamefont {Tung}},
  \bibinfo {author} {\bibfnamefont {B.~G.}\ \bibnamefont {Sumpter}}, \bibinfo
  {author} {\bibfnamefont {W.-R.}\ \bibnamefont {Chen}},\ and\ \bibinfo
  {author} {\bibfnamefont {C.}~\bibnamefont {Do}},\ }\bibfield  {title}
  {\enquote {\bibinfo {title} {Deciphering the scattering of mechanically
  driven polymers using deep learning},}\ }\href@noop {} {\bibfield  {journal}
  {\bibinfo  {journal} {Journal of Chemical Theory and Computation}\ }\textbf
  {\bibinfo {volume} {21}},\ \bibinfo {pages} {4176--4182} (\bibinfo {year}
  {2025}{\natexlab{c}})}\BibitemShut {NoStop}%
\bibitem [{\citenamefont {Cranmer}, \citenamefont {Brehmer},\ and\
  \citenamefont {Louppe}(2020)}]{cranmer2020frontier}%
  \BibitemOpen
  \bibfield  {author} {\bibinfo {author} {\bibfnamefont {K.}~\bibnamefont
  {Cranmer}}, \bibinfo {author} {\bibfnamefont {J.}~\bibnamefont {Brehmer}},\
  and\ \bibinfo {author} {\bibfnamefont {G.}~\bibnamefont {Louppe}},\
  }\bibfield  {title} {\enquote {\bibinfo {title} {The frontier of
  simulation-based inference},}\ }\href@noop {} {\bibfield  {journal} {\bibinfo
   {journal} {Proceedings of the National Academy of Sciences}\ }\textbf
  {\bibinfo {volume} {117}},\ \bibinfo {pages} {30055--30062} (\bibinfo {year}
  {2020})}\BibitemShut {NoStop}%
\bibitem [{\citenamefont {Kingma}, \citenamefont {Welling}\ \emph
  {et~al.}(2013)\citenamefont {Kingma}, \citenamefont {Welling} \emph
  {et~al.}}]{kingma2013auto}%
  \BibitemOpen
  \bibfield  {author} {\bibinfo {author} {\bibfnamefont {D.~P.}\ \bibnamefont
  {Kingma}}, \bibinfo {author} {\bibfnamefont {M.}~\bibnamefont {Welling}},
  \emph {et~al.},\ }\href@noop {} {\enquote {\bibinfo {title} {Auto-encoding
  variational bayes},}\ } (\bibinfo {year} {2013})\BibitemShut {NoStop}%
\bibitem [{\citenamefont {Doersch}(2016)}]{doersch2016tutorial}%
  \BibitemOpen
  \bibfield  {author} {\bibinfo {author} {\bibfnamefont {C.}~\bibnamefont
  {Doersch}},\ }\bibfield  {title} {\enquote {\bibinfo {title} {Tutorial on
  variational autoencoders},}\ }\href@noop {} {\bibfield  {journal} {\bibinfo
  {journal} {arXiv preprint arXiv:1606.05908}\ } (\bibinfo {year}
  {2016})}\BibitemShut {NoStop}%
\bibitem [{\citenamefont {Allen}\ \emph {et~al.}(2004)\citenamefont {Allen}
  \emph {et~al.}}]{allen2004introduction}%
  \BibitemOpen
  \bibfield  {author} {\bibinfo {author} {\bibfnamefont {M.~P.}\ \bibnamefont
  {Allen}} \emph {et~al.},\ }\bibfield  {title} {\enquote {\bibinfo {title}
  {Introduction to molecular dynamics simulation},}\ }\href@noop {} {\bibfield
  {journal} {\bibinfo  {journal} {Computational soft matter: from synthetic
  polymers to proteins}\ }\textbf {\bibinfo {volume} {23}},\ \bibinfo {pages}
  {1--28} (\bibinfo {year} {2004})}\BibitemShut {NoStop}%
\bibitem [{\citenamefont {Thompson}\ \emph {et~al.}(2022)\citenamefont
  {Thompson}, \citenamefont {Aktulga}, \citenamefont {Berger}, \citenamefont
  {Bolintineanu}, \citenamefont {Brown}, \citenamefont {Crozier}, \citenamefont
  {In't~Veld}, \citenamefont {Kohlmeyer}, \citenamefont {Moore}, \citenamefont
  {Nguyen} \emph {et~al.}}]{thompson2022lammps}%
  \BibitemOpen
  \bibfield  {author} {\bibinfo {author} {\bibfnamefont {A.~P.}\ \bibnamefont
  {Thompson}}, \bibinfo {author} {\bibfnamefont {H.~M.}\ \bibnamefont
  {Aktulga}}, \bibinfo {author} {\bibfnamefont {R.}~\bibnamefont {Berger}},
  \bibinfo {author} {\bibfnamefont {D.~S.}\ \bibnamefont {Bolintineanu}},
  \bibinfo {author} {\bibfnamefont {W.~M.}\ \bibnamefont {Brown}}, \bibinfo
  {author} {\bibfnamefont {P.~S.}\ \bibnamefont {Crozier}}, \bibinfo {author}
  {\bibfnamefont {P.~J.}\ \bibnamefont {In't~Veld}}, \bibinfo {author}
  {\bibfnamefont {A.}~\bibnamefont {Kohlmeyer}}, \bibinfo {author}
  {\bibfnamefont {S.~G.}\ \bibnamefont {Moore}}, \bibinfo {author}
  {\bibfnamefont {T.~D.}\ \bibnamefont {Nguyen}}, \emph {et~al.},\ }\bibfield
  {title} {\enquote {\bibinfo {title} {Lammps-a flexible simulation tool for
  particle-based materials modeling at the atomic, meso, and continuum
  scales},}\ }\href@noop {} {\bibfield  {journal} {\bibinfo  {journal}
  {Computer physics communications}\ }\textbf {\bibinfo {volume} {271}},\
  \bibinfo {pages} {108171} (\bibinfo {year} {2022})}\BibitemShut {NoStop}%
\bibitem [{\citenamefont {Lennard-Jones}(1931)}]{lennard1931cohesion}%
  \BibitemOpen
  \bibfield  {author} {\bibinfo {author} {\bibfnamefont {J.~E.}\ \bibnamefont
  {Lennard-Jones}},\ }\bibfield  {title} {\enquote {\bibinfo {title}
  {Cohesion},}\ }\href@noop {} {\bibfield  {journal} {\bibinfo  {journal}
  {Proceedings of the Physical Society}\ }\textbf {\bibinfo {volume} {43}},\
  \bibinfo {pages} {461} (\bibinfo {year} {1931})}\BibitemShut {NoStop}%
\bibitem [{\citenamefont {Nos{\'e}}(1984)}]{nose1984unified}%
  \BibitemOpen
  \bibfield  {author} {\bibinfo {author} {\bibfnamefont {S.}~\bibnamefont
  {Nos{\'e}}},\ }\bibfield  {title} {\enquote {\bibinfo {title} {A unified
  formulation of the constant temperature molecular dynamics methods},}\
  }\href@noop {} {\bibfield  {journal} {\bibinfo  {journal} {The Journal of
  chemical physics}\ }\textbf {\bibinfo {volume} {81}},\ \bibinfo {pages}
  {511--519} (\bibinfo {year} {1984})}\BibitemShut {NoStop}%
\bibitem [{\citenamefont {Hoover}(1985)}]{hoover1985canonical}%
  \BibitemOpen
  \bibfield  {author} {\bibinfo {author} {\bibfnamefont {W.~G.}\ \bibnamefont
  {Hoover}},\ }\bibfield  {title} {\enquote {\bibinfo {title} {Canonical
  dynamics: Equilibrium phase-space distributions},}\ }\href@noop {} {\bibfield
   {journal} {\bibinfo  {journal} {Physical review A}\ }\textbf {\bibinfo
  {volume} {31}},\ \bibinfo {pages} {1695} (\bibinfo {year}
  {1985})}\BibitemShut {NoStop}%
\bibitem [{\citenamefont {Guinier}\ \emph {et~al.}(1956)\citenamefont
  {Guinier}, \citenamefont {Fournet}, \citenamefont {Walker},\ and\
  \citenamefont {Vineyard}}]{guinier1956small}%
  \BibitemOpen
  \bibfield  {author} {\bibinfo {author} {\bibfnamefont {A.}~\bibnamefont
  {Guinier}}, \bibinfo {author} {\bibfnamefont {G.}~\bibnamefont {Fournet}},
  \bibinfo {author} {\bibfnamefont {C.~B.}\ \bibnamefont {Walker}},\ and\
  \bibinfo {author} {\bibfnamefont {G.~H.}\ \bibnamefont {Vineyard}},\
  }\href@noop {} {\enquote {\bibinfo {title} {Small-angle scattering of
  x-rays},}\ } (\bibinfo {year} {1956})\BibitemShut {NoStop}%
\bibitem [{\citenamefont {Ornstein}(1914)}]{ornstein1914accidental}%
  \BibitemOpen
  \bibfield  {author} {\bibinfo {author} {\bibfnamefont {L.~S.}\ \bibnamefont
  {Ornstein}},\ }\bibfield  {title} {\enquote {\bibinfo {title} {Accidental
  deviations of density and opalescence at the critical point of a single
  substance},}\ }\href@noop {} {\bibfield  {journal} {\bibinfo  {journal}
  {Proc. Akad. Sci.}\ }\textbf {\bibinfo {volume} {17}},\ \bibinfo {pages}
  {793} (\bibinfo {year} {1914})}\BibitemShut {NoStop}%
\bibitem [{\citenamefont {Kinning}\ and\ \citenamefont
  {Thomas}(1984)}]{kinning1984hard}%
  \BibitemOpen
  \bibfield  {author} {\bibinfo {author} {\bibfnamefont {D.~J.}\ \bibnamefont
  {Kinning}}\ and\ \bibinfo {author} {\bibfnamefont {E.~L.}\ \bibnamefont
  {Thomas}},\ }\bibfield  {title} {\enquote {\bibinfo {title} {Hard-sphere
  interactions between spherical domains in diblock copolymers},}\ }\href@noop
  {} {\bibfield  {journal} {\bibinfo  {journal} {Macromolecules}\ }\textbf
  {\bibinfo {volume} {17}},\ \bibinfo {pages} {1712--1718} (\bibinfo {year}
  {1984})}\BibitemShut {NoStop}%
\bibitem [{\citenamefont {Scheffold}(2025)}]{SCHEFFOLD2025581}%
  \BibitemOpen
  \bibfield  {author} {\bibinfo {author} {\bibfnamefont {F.}~\bibnamefont
  {Scheffold}},\ }\bibfield  {title} {\enquote {\bibinfo {title} {Chapter 21 -
  light scattering and propagation in turbid media},}\ }in\ \href
  {https://doi.org/https://doi.org/10.1016/B978-0-443-29116-6.00008-4} {\emph
  {\bibinfo {booktitle} {Neutrons, X-rays, and Light (Second Edition)}}},\
  \bibinfo {editor} {edited by\ \bibinfo {editor} {\bibfnamefont
  {P.}~\bibnamefont {Lindner}}\ and\ \bibinfo {editor} {\bibfnamefont
  {J.}~\bibnamefont {Oberdisse}}}\ (\bibinfo  {publisher} {Elsevier},\ \bibinfo
  {year} {2025})\ \bibinfo {edition} {second edition}\ ed.,\ pp.\ \bibinfo
  {pages} {581--614}\BibitemShut {NoStop}%
\bibitem [{\citenamefont {Paszke}(2019)}]{paszke2019pytorch}%
  \BibitemOpen
  \bibfield  {author} {\bibinfo {author} {\bibfnamefont {A.}~\bibnamefont
  {Paszke}},\ }\bibfield  {title} {\enquote {\bibinfo {title} {Pytorch: An
  imperative style, high-performance deep learning library},}\ }\href@noop {}
  {\bibfield  {journal} {\bibinfo  {journal} {arXiv preprint arXiv:1912.01703}\
  } (\bibinfo {year} {2019})}\BibitemShut {NoStop}%
\bibitem [{\citenamefont {Kingma}(2014)}]{kingma2014adam}%
  \BibitemOpen
  \bibfield  {author} {\bibinfo {author} {\bibfnamefont {D.~P.}\ \bibnamefont
  {Kingma}},\ }\bibfield  {title} {\enquote {\bibinfo {title} {Adam: A method
  for stochastic optimization},}\ }\href@noop {} {\bibfield  {journal}
  {\bibinfo  {journal} {arXiv preprint arXiv:1412.6980}\ } (\bibinfo {year}
  {2014})}\BibitemShut {NoStop}%
\bibitem [{\citenamefont {Loshchilov}\ and\ \citenamefont
  {Hutter}(2016)}]{loshchilov2016sgdr}%
  \BibitemOpen
  \bibfield  {author} {\bibinfo {author} {\bibfnamefont {I.}~\bibnamefont
  {Loshchilov}}\ and\ \bibinfo {author} {\bibfnamefont {F.}~\bibnamefont
  {Hutter}},\ }\bibfield  {title} {\enquote {\bibinfo {title} {Sgdr: Stochastic
  gradient descent with warm restarts},}\ }\href@noop {} {\bibfield  {journal}
  {\bibinfo  {journal} {arXiv preprint arXiv:1608.03983}\ } (\bibinfo {year}
  {2016})}\BibitemShut {NoStop}%
\bibitem [{\citenamefont {Jolliffe}\ and\ \citenamefont
  {Cadima}(2016)}]{jolliffe2016principal}%
  \BibitemOpen
  \bibfield  {author} {\bibinfo {author} {\bibfnamefont {I.~T.}\ \bibnamefont
  {Jolliffe}}\ and\ \bibinfo {author} {\bibfnamefont {J.}~\bibnamefont
  {Cadima}},\ }\bibfield  {title} {\enquote {\bibinfo {title} {Principal
  component analysis: a review and recent developments},}\ }\href@noop {}
  {\bibfield  {journal} {\bibinfo  {journal} {Philosophical transactions of the
  royal society A: Mathematical, Physical and Engineering Sciences}\ }\textbf
  {\bibinfo {volume} {374}},\ \bibinfo {pages} {20150202} (\bibinfo {year}
  {2016})}\BibitemShut {NoStop}%
\bibitem [{\citenamefont {Golub}\ and\ \citenamefont
  {Kahan}(1965)}]{golub1965calculating}%
  \BibitemOpen
  \bibfield  {author} {\bibinfo {author} {\bibfnamefont {G.}~\bibnamefont
  {Golub}}\ and\ \bibinfo {author} {\bibfnamefont {W.}~\bibnamefont {Kahan}},\
  }\bibfield  {title} {\enquote {\bibinfo {title} {Calculating the singular
  values and pseudo-inverse of a matrix},}\ }\href@noop {} {\bibfield
  {journal} {\bibinfo  {journal} {Journal of the Society for Industrial and
  Applied Mathematics, Series B: Numerical Analysis}\ }\textbf {\bibinfo
  {volume} {2}},\ \bibinfo {pages} {205--224} (\bibinfo {year}
  {1965})}\BibitemShut {NoStop}%
\bibitem [{\citenamefont {Imhof}\ and\ \citenamefont
  {Dhont}(1995)}]{imhof1995experimental}%
  \BibitemOpen
  \bibfield  {author} {\bibinfo {author} {\bibfnamefont {A.}~\bibnamefont
  {Imhof}}\ and\ \bibinfo {author} {\bibfnamefont {J.}~\bibnamefont {Dhont}},\
  }\bibfield  {title} {\enquote {\bibinfo {title} {Experimental phase diagram
  of a binary colloidal hard-sphere mixture with a large size ratio},}\
  }\href@noop {} {\bibfield  {journal} {\bibinfo  {journal} {Physical review
  letters}\ }\textbf {\bibinfo {volume} {75}},\ \bibinfo {pages} {1662}
  (\bibinfo {year} {1995})}\BibitemShut {NoStop}%
\bibitem [{\citenamefont {Schneider}\ \emph {et~al.}(1987)\citenamefont
  {Schneider}, \citenamefont {Karrer}, \citenamefont {Dhont},\ and\
  \citenamefont {Klein}}]{schneider1987pair}%
  \BibitemOpen
  \bibfield  {author} {\bibinfo {author} {\bibfnamefont {J.}~\bibnamefont
  {Schneider}}, \bibinfo {author} {\bibfnamefont {D.}~\bibnamefont {Karrer}},
  \bibinfo {author} {\bibfnamefont {J.}~\bibnamefont {Dhont}},\ and\ \bibinfo
  {author} {\bibfnamefont {R.}~\bibnamefont {Klein}},\ }\bibfield  {title}
  {\enquote {\bibinfo {title} {The pair-distribution function and
  light-scattered intensities for charged rod-like macromolecules in
  solution},}\ }\href@noop {} {\bibfield  {journal} {\bibinfo  {journal} {The
  Journal of chemical physics}\ }\textbf {\bibinfo {volume} {87}},\ \bibinfo
  {pages} {3008--3015} (\bibinfo {year} {1987})}\BibitemShut {NoStop}%
\end{thebibliography}%

\end{document}